\documentclass[a4paper,fleqn,usenatbib]{mnras}

\usepackage{newtxtext,newtxmath}

\usepackage[T1]{fontenc}
\usepackage{ae,aecompl}

\usepackage{graphicx}	
\usepackage{amsmath}	
\usepackage{amssymb}	
\usepackage{verbatim} 
\usepackage{subcaption}
\def \Msunh{\ h^{-1}{\rm M_\odot}}

\def \Mpch{\ h^{-1}{\rm Mpc}}

\def \lcdm{$\Lambda$CDM}
\def \qcdm{$\phi$CDM}

\newcommand{\Tableref}[1]{Table~\ref{#1}}

\newcommand{\ncosmo}{3}

\title[Cosmic Voids in Evolving Dark Sector Cosmologies]{Cosmic Voids in Evolving Dark Sector Cosmologies: the Low Redshift Universe}

\author[E. Adermann et al.]{
Eromanga Adermann,$^{1}$\thanks{E-mail: eromanga.adermann@sydney.edu.au (EA)}
Pascal J. Elahi,$^{1,2}$
Geraint F. Lewis$^{1}$
and Chris Power$^{2}$
\\
$^{1}$Sydney Institute for Astronomy, School of Physics, A28, The University of Sydney, NSW, 2006, Australia\\
$^{2}$International Centre for Radio Astronomy Research (ICRAR), The University of Western Australia, 35 Stirling Hwy, \\
Crawley, Western Australia 6009, Australia}

\date{Accepted XXX. Received YYY; in original form ZZZ}

\pubyear{2016}

\begin{document}
\label{firstpage}
\pagerange{\pageref{firstpage}--\pageref{lastpage}}
\maketitle

\begin{abstract}
We present a comparison of void properties between the standard model of cosmology, $\Lambda$ Cold Dark Matter ($\Lambda$CDM), and two alternative cosmological models with evolving and interacting dark sectors: a quintessence model ($\phi$CDM) and a Coupled Dark Matter-Dark Energy (CDE) model. Using $N$-body simulations of these models, we derive several measures of void statistics and properties, including distributions of void volume, ellipticity, prolateness, and average density. We find that the volume distribution derived from the CDE simulation deviates from the volume distribution derived from the $\Lambda$CDM simulation in the present-day universe, suggesting that the presence of a coupled dark sector could be observable through this statistic. We also find that the distributions of void ellipticity and prolateness are practically indistinguishable among the three models over the redshift range $z=0.0-1.0$, indicating that simple void shape statistics are insensitive to small changes in dark sector physics. Interestingly, we find that the distributions of average void density measured in each of the three simulations are distinct from each other. In particular, voids on average tend to be emptiest under a quintessence model, and densest under the $\Lambda$CDM model. Our results suggest that it is the scalar field present in both alternative models that causes emptier voids to form, while the coupling of the dark sector mitigates this effect by slowing down the evacuation of matter from voids.  

\end{abstract}

\begin{keywords}
cosmology: simulations -- cosmology: large-scale structure of the universe -- voids -- dark matter -- dark energy
\end{keywords}



\section{Introduction}

Cosmic voids have recently gained popularity as cosmological probes \citep[e.g.][]{2014MNRAS.445.1235S, 2015BAAA...57....6P}, primarily because these large, underdense regions of the universe are only mildly non-linear, experiencing little gravitational evolution compared to other, denser structures that make up the cosmic web. The statistics of these denser structures, namely sheets, filaments and knots (that house dark matter haloes), are frequently used as probes of cosmology, for example in the use of the halo mass function \citep{2008ApJ...687....7S} and the properties of galaxy cluster gas \citep{2010MNRAS.403.1684B, 2014MNRAS.439.2958C}. The difficulty with this approach is that the properties and evolution of overdense regions are typically governed by more than just underlying cosmology. Complex baryonic physics and coupling due to non-linear evolution also play a significant role in governing these structures, which at best complicates the process of distinguishing between baryonic physics and effects due to cosmology, and at worst may mask the cosmological signature we are searching for.  In contrast, baryonic physics leaves a much smaller imprint on the growth of cosmic voids. For this reason, voids are seen as relatively pristine environments ideal for studying cosmology. 

Voids have been used in a variety of cosmological studies, due to the low density environment within them which makes them particularly sensitive probes of cosmology. These include studies of the integrated Sachs-Wolfe effect \citep[e.g.][]{2008ApJ...683L..99G, 2014ApJ...786..110C, 2015MNRAS.454.2804G} and weak gravitational lensing studies \citep[e.g.][]{2015MNRAS.454.3357C}. Voids have also been established as useful probes of dark energy \citep[e.g.][]{2012MNRAS.426..440B, 2015PhRvD..92h3531P}, modified gravity theories \citep[e.g.][]{2013MNRAS.431..749C, 2015MNRAS.451.4215Z, 2015MNRAS.451.1036C, 2015MNRAS.450.3319L, 2015JCAP...08..028B, 2016arXiv160901284A}, and alternative cosmological models \citep[e.g.][]{2015MNRAS.446L...1S, 2015MNRAS.451.3606Y, 2015JCAP...11..018M, 2016MNRAS.455.3075P}, the last of which will be the focus of this paper. 

The $\Lambda$ Cold Dark Matter ($\Lambda$CDM) model of cosmology is the standard model for the universe. Within this cosmology, the universe is spatially flat, and dominated by an invisible dark sector composed of dark matter and dark energy. Dark matter plays a significant role in the clustering of baryonic matter and the formation of structure in the universe, while a cosmological constant $\Lambda$, or dark energy, drives the late-time accelerated expansion of the universe \citep{1998AJ....116.1009R, 1999ApJ...517..565P}. 

The $\Lambda$CDM cosmology is well-supported by observational results. For example, the anisotropies in the Cosmic Microwave Background (CMB; e.g. \citealt{2013ApJS..208...20B}; \citealt{2014A&A...571A..23P}; \citealt{2015arXiv150607135P}),  features in the Large-Scale Structure (e.g. \citealt{2009ApJS..182..543A}; \citealt{2012MNRAS.423.3430B}), Baryonic Acoustic Oscillations (\citealt{2011MNRAS.416.3017B}; \citealt{2011MNRAS.418.1707B}; \citealt{2014MNRAS.441...24A}), weak lensing (e.g. \citealt{2013MNRAS.430.2200K}; \citealt{2013MNRAS.432.2433H}), cluster abundances (e.g.\citealt{2009ApJ...692.1060V}; \citealt{2010ApJ...708..645R}), galaxy clustering (e.g. \citealt{2004PhRvD..69j3501T}; \citealt{2010MNRAS.404...60R}), and the luminosity-distance relation from Type Ia supernovae (e.g. \citealt{2008ApJ...686..749K}; \citealt{2011ApJS..192....1C}; \citealt{2012ApJ...746...85S}) are all consistent with $\Lambda$CDM.  

Although $\Lambda$CDM offers a framework for dark matter and dark energy, it does very little to explain or predict their existence, properties and behaviour. Instead, we must rely heavily on observations to pinpoint the properties of the dark sector. Such observations have supported `cold' dark matter \citep[which has well-motivated candidates from theoretical particle physics; see][]{2005PhR...405..279B, 2013IJMPA..2830028P} over more energetic dark matter (such as hot dark matter which is relativistic at decoupling, e.g. \citealt{1994ApJ...434L...5M}; \citealt{1993MNRAS.265..507P}). There is strong evidence for (a) non-baryonic elementary dark matter particle(s), though the exact properties of such (a) particle(s) are poorly constrained by observations.  Dark energy appears to be well-characterised by a cosmological constant, $\Lambda$, driving the late time accelerated expansion and with a measured equation of state given by $w = \rho/p \approx -1$ \citep{2012ApJ...746...85S, 2016MNRAS.461.3781C}. However, its origin and nature is not well-characterised. Without a more in-depth framework for the dark sector whose predictions match observations as well as the standard model, the nature of the dark sector remains an unsolved problem. 

In spite of the excellent consistency between the predictions of $\Lambda$CDM and observations of the large-scale structure, the standard model does not appear to be in agreement with observations on galaxy scales. For example, $\Lambda$CDM predicts many more satellite galaxies than are observed for galaxies like the Milky Way (\citealt{1999ApJ...522...82K}; \citealt{1999ApJ...524L..19M}). While this may be the result of feedback processes removing gas from low mass subhaloes and leaving almost completely dark subhaloes \citep[e.g.][] {2000ApJ...539..517B, 2002MNRAS.333..177B, 2011MNRAS.415..257N, 2012ASPC..453..305N, 2014MNRAS.445..581H, 2016MNRAS.457.1931S}, the missing satellites may also indicate the need for more energetic dark matter candidates, or Warm Dark Matter (e.g. \citealt{2013PASA...30...53P}; \citealt{2014MNRAS.444.2333E}). Another inconsistency is the unexplained alignment of satellite galaxies around the Milky Way  (known as the Vast Polar Structure; \citealt{2012MNRAS.423.1109P}) and Andromeda (known as the Plane of Satellites; \citealt{2013Natur.493...62I}; \citealt{2013ApJ...766..120C}). Such alignments have also been seen beyond the local group (e.g. \citealt{2006MNRAS.369.1293Y}; \citealt{2013ApJ...768...20L}). Furthermore, the velocities of these satellite galaxies are significantly more correlated than $\Lambda$CDM predicts (e.g. \citealt{2014Natur.511..563I}; \citealt{2015ApJ...800...34G}), though there is some evidence that these alignments are not too improbable in $\Lambda$CDM \citep{2015MNRAS.452.3838C, 2016MNRAS.457.1931S}. 

Within the $\Lambda$CDM framework, the densities of dark matter and dark energy evolve independently, as different functions of cosmic time. Dark matter dominated at early times, allowing structure to form, while dark energy dominates the current epoch, driving the accelerated expansion of the universe. However, in spite of their independence, the two densities coincidentally have similar values today. This is dubbed the coincidence problem, and its existence suggests that there may be some inter-dependence between dark matter and dark energy, which is currently absent from $\Lambda$CDM (e.g. \citealt{1995A&A...301..321W}; \citealt{2000PhRvD..62d3511A}). This has led to the proposal of alternative models, including models with modifications to general relativity (e.g. \citealt{2007PhRvD..76f4004H}; \citealt{1980PhLB...91...99S}; see \citealt{2010LRR....13....3D} for a review), and models with dynamical scalar fields, or quintessence models (e.g. \citealt{1988PhRvD..37.3406R}; \citealt{1988NuPhB.302..668W}; \citealt{2001PhRvD..63j3510A}; see \citealt{2013CQGra..30u4003T} for a review). 

In dynamical scalar field models, dark energy is assumed to couple to dark matter (regardless of whether the dark matter is cold or warm). Dark matter decaying into the scalar dark energy field results in the late-time accelerated expansion of the Universe. The coupled dark sector gives rise to dark matter particles that change with time, and the strength of gravitational interactions between dark matter particles differs from the corresponding strength of baryon-baryon and dark matter-baryon interactions.  Former studies into structure formation in coupled versus uncoupled models have revealed significant differences in the growth functions, and thus the matter power spectrum, between the models (\citealt{2011PhRvD..83b4007L}). Differences have also been found in the weak lensing signature of coupled models (\citealt{2015MNRAS.447..858P}, \citealt{2015MNRAS.452.2757G}). \citealt{2016MNRAS.455.3075P} found that coupled dark energy models have a greater number of large voids in the cold dark matter distribution than $\Lambda$CDM, which they suggest is the result of a higher normalisation of linear perturbations at low redshifts in the coupled models.  \citet{2015MNRAS.446L...1S} found similar differences in the number of very large voids between a strongly coupled model (ruled out by observations; \citealt{2012PhRvD..86j3507P}) and $\Lambda$CDM, and additionally found differences in the void density profiles between these cosmologies. However, \citet{2014MNRAS.439.2943C, 2014MNRAS.439.2958C}  found no significant changes in the cosmic web and halo mass function in coupled models, although they did discover small differences in the concentration and spin parameter of small field haloes. 

The goal of our study is to further probe the signatures of alternative cosmological models. We examine void properties in the low redshift Universe, determine how they differ among cosmological models and propose explanations for the signatures found. Further investigations into void properties and their evolution in a high redshift universe will be published in later papers. In Sections \ref{SiriusBlack} and \ref{Hewhoshallnotbenamed}, we present the simulations of the standard and alternative cosmological models, as well as the methods we used to identify voids in the large-scale structure. Our main results are presented in Section \ref{Luna}, followed by a discussion and explanation of the results in Section \ref{Bellatrix}. 

\section{Evolving Dark Sector Models}\label{SiriusBlack}
We focus on two non-standard cosmological models, both of which feature dynamic dark sectors: an uncoupled quintessence model (\qcdm) and a coupled dark-energy dark matter model, which will be referred to as Coupled Dark Energy (CDE) throughout this paper. Both these models differ from $\Lambda$CDM in that dark energy arises due to the evolution of a time-dependent scalar field, $\phi$, rather than being characterised as a cosmological constant. Additionally, the scalar field may be coupled to dark matter in general evolving dark sector models. The general form of the Lagrangian for the scalar field of such models (including the two models of interest) is given by:  

\begin{equation}
L = \int\sqrt{-g} \left(-\frac{1}{2}\partial_{\mu}\partial^{\mu}\phi + V(\phi) + m(\phi)\psi_{m}\bar{\psi}_{m}\right)d^{4}x,
\end{equation} 
consisting of a kinetic term, a potential term $V(\phi)$ which if chosen appropriately gives rise to the late-time accelerated expansion, and a coupling term that describes the interaction between the scalar field and dark matter field ($\psi_{m}$). For our models, we choose a Ratra-Peebles potential \citep{1988PhRvD..37.3406R}:  
\begin{equation}
V(\phi) = V_{0}\phi^{-\alpha}
\end{equation} 
where $V_{0}$ and $\alpha$ are constants, determined through fits to observational data, and $\phi$ is in Planck units. 

\subsection{Uncoupled Quintessence  (\qcdm)}
The Lagrangian for the scalar field of the uncoupled quintessence model contains no coupling term ($m(\phi)=0$), and hence there is no direct interaction between the scalar field $\phi$ and the dark matter field $\psi_{m}$. However, unlike \lcdm, the dark energy density evolves with time as a result of the inclusion of the scalar field, which in turn alters the expansion history of the universe compared to \lcdm. 
\subsection{Coupled Dark Energy (CDE)} 
The coupling mechanism between the scalar field and the dark matter field is enabled by choosing a non-zero form of $m(\phi)$, also known as the interaction term. For our CDE model, we choose 
\begin{equation}
m(\phi) = m_{0} \mathrm{exp}[-\beta(\phi)\phi], 
\end{equation}
for the interaction term, a form that alleviates the coincidence problem. To simplify things, we choose a constant coupling term, $\beta(\phi) = \beta_{o}=0.05$. This coupling enables dark matter particles to decay into the scalar field. It also results in an additional frictional force experienced by dark matter particles (but not baryons), leading to a difference in the evolution of the amplitude of their density perturbations. This coupling has been found to significantly affect the fraction of baryons in cluster-sized haloes (\citealt{2010MNRAS.403.1684B}). Our CDE model differs from $\Lambda$CDM in both the expansion history, due to the scalar field, and the evolution of the density perturbation amplitudes, due to the coupling. 

\section{Simulations} \label{Hewhoshallnotbenamed} 
For our study, we produced simulations of the following \ncosmo\ cosmologies: a reference \lcdm, an uncoupled quintessence model, and a coupled dark energy-dark matter model with a coupling parameter $\beta_o$. Following \cite{2015MNRAS.452.1341E}, we choose $(h,\Omega_m,\Omega_b,\sigma_8)=(0.67,0.3175,0.049,0.83)$, ensuring all cosmologies have parameters consistent with $z=0$ \lcdm\ Planck data  \cite[][]{2014A&A...571A..16P, 2016A&A...594A..13P}. The coupling parameter, $\beta_o=0.05$, in our simulation is chosen to test the boundaries of allowed coupling \cite[see][]{2012PhRvD..86j3507P, 2013JCAP...11..022X} in order to maximise any observational differences that may exist between this cosmology and the standard \lcdm\ model. The linear power spectrum and growth factor $f\equiv d\ln D(a)/d\ln a$ are calculated using first-order Newtonian perturbation equations and the publicly available Boltzmann code {\sc cmbeasy} \cite[][]{2005JCAP...10..011D}. We note that although these simulations have neglected star formation and feedback physics, baryons do not appear to significantly change the void population. Large underdense regions remain underdense, regardless of whether the simulation includes dark matter only, or accounts for hydrodynamic or full baryonic physics \citep{2016arXiv160900101P}. 

\par 
Initial conditions are generated by perturbing particles placed on a Cartesian grid with the first-order Zel'Dovich approximation using a modified version of the publicly available {\sc n-genic} code. The modified code uses the growth factors calculated by {\sc cmbeasy} to correctly calculate the particle displacements in the non-standard cosmologies. All the simulations are started at $z=100$ with the same phases in the density perturbations. Choosing the same initial phases results in underdense regions forming in the same locations in all simulations. Each underdense region in a given simulation has a corresponding underdense region in other simulations (perhaps with different properties such as density or size, depending on cosmology). This is the case despite potentially different exact profiles of density perturbations between cosmologies with different power spectra, such as $\Lambda$CDM and CDE. 

\par
We use {\sc dark-gadget}, a modified version of {\sc p-gadget}-2 N-Body code \cite[for more details see][]{2014MNRAS.439.2943C}. The key modifications are the inclusion of a separate gravity tree to account for the additional long range forces arising from the scalar field, and an evolving dark matter N-body particle mass which models the decay of the dark matter density. The code requires the full evolution of the scalar field, $\phi$, the mass of the dark matter N-body particle, and the expansion history. 

\par
\begin{figure*}
\includegraphics[width=\linewidth]{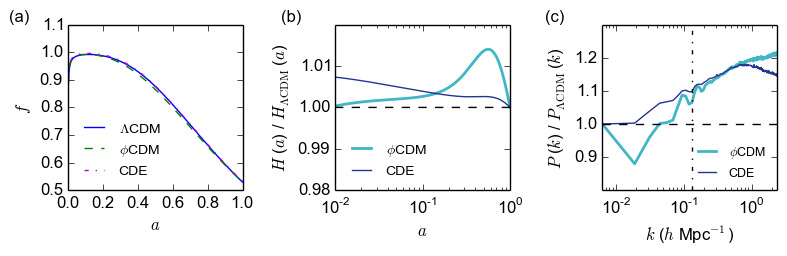}
\caption{(a) The evolution of the growth factor, $f$, as a function of the scale factor, $a$, for the $\Lambda$CDM (solid blue), $\phi$CDM (dashed green) and CDE (dot-dashed magenta) simulations. (b) The ratio of the Hubble constant in the $\phi$CDM (thick teal) and CDE (thin blue) simulations, $H(a)$, to the Hubble constant in the $\Lambda$CDM simulation, $H_{\Lambda \mathrm{CDM}}(a)$, expressed as a function of the scale factor $a$. (c) The ratio of the non-linear power spectrum in the $\phi$CDM (thick teal) and CDE (thin blue) simulations, $P(k)$, to the power spectrum in the $\Lambda$CDM simulation, $P_{\Lambda \mathrm{CDM}}(k)$, expressed as a function of the wavenumber $k$. To the left of the vertical line is the region where the ratios have an error of less than $1\%$. }
\label{Sims} 
\end{figure*}

Our simulations are $500\Mpch$ boxes containing $2\times512^3$ particles (dark matter and gas) giving a mass resolution of $m_{\rm dm} (m_{\rm gas}) = 6.9 (1.3)\times10^{10}\Msunh$\footnote{The exact masses in the coupled cosmology naturally depend on redshift, but this is the mass resolution at $z=0$, as we fix $\Omega_m$ at this time.}. Though the mass resolution is poor for identifying haloes, this resolution is more than enough to identify voids. All runs use a gravitational softening length of $1/33$ of the interparticle spacing. The set of simulations run is summarised in \Tableref{tab:sims}. Figure~\ref{Sims} shows the growth factor evolution, the evolution of the Hubble constant and the non-linear power spectrum in the three simulations.   
\begin{table}
\setlength\tabcolsep{2pt}
\centering\footnotesize
\caption{Simulations}
\begin{tabular}{@{\extracolsep{\fill}}l p{0.35\textwidth}}
\hline
\hline
    Cosmology & Comments\\
\hline
    \lcdm       & DM/gas run along with 5 pure DM runs $(m_{\rm DM}=8.1\times10^{10}\Msunh)$ each with different random seeds to study cosmic variance. \\
    \qcdm       & DM/gas run along with 2 pure DM runs each with different random seeds to study cosmic variance. \\
    CDE & DM/gas run only, with $\beta_o = 0.05$. \\
\hline
\end{tabular}
\label{tab:sims}
\end{table}

%
%
%

\section{Void Finding} \label{Luna}
Our study focuses on voids in the cold dark matter particle distributions of our simulations. The method we used to define and identify these voids is based on the Hessian matrix of second derivatives at each point, a method which has also been used to identify other large-scale structures, such as sheets, filaments and knots \citep[e.g.][]{2007MNRAS.375..489H, 2007ApJ...655L...5A, 2009ApJ...706..747Z, 2013ApJ...762...72T}. The Hessian matrix is given by
\begin{equation}\label{Hessian}
H_{\alpha\beta} (\mathbf x) = \frac{\partial^{2} \rho(\mathbf x)}{\partial x_{\alpha} \partial x_{\beta}}, 
\end{equation}
which characterises the curvature of a density field, $\rho(\textbf{x})$, as a function of position, $\textbf{x}$. The eigenvalues of this matrix reveal whether the local density field around the point $\textbf{x}$ is a local maximum or minimum, along three orthogonal directions (given by its eigenvectors). If all three eigenvalues are negative, the local density field is a local maximum, or is knot-like. If only two eigenvalues are negative, the density field has a local maximum along two orthogonal axes and a local minimum along the third axis, or is filament-like. If one eigenvalue is negative, then the density field has a local maximum along only one of its axes, or is sheet-like. Finally, if all three eigenvalues are positive, the density field exhibits a local minimum in all directions, or is void-like. 

To find the void regions in the dark matter density field, we divided the simulation up into a regular grid of $500 \times 500 \times 500$ cells, so that each cell was 1 $h^{-1}$Mpc in length. The densities of particles were estimated using a smooth particle hydrodynamics kernel. These densities were then assigned to the nearest grid cell, so that an average density for each cell could be calculated. Finally, these density values were convolved with a Gaussian kernel, with a scale of 3 $h^{-1}$Mpc, to smooth out small-scale fluctuations. Both the cell length and smoothing scale were chosen to match the visual classification of structure based on particle positions only. We calculated the eigenvalues of the Hessian matrix for each cell, and gave each of them a classification using the scheme described above. We then utilised a friends-of-friends algorithm to group neighbouring void-like cells together, requiring that neighbouring cells be linked together only if both are void-like cells, or the second is a sheet-like cell. This results in a single layer of sheet-like cells surrounding and bounding each group of linked void-like cells, naturally defining a physical surface with a thickness of 1 $h^{-1}$Mpc for our voids. We also imposed the requirement that each void consists of at least two void-like cells linked together, with a corresponding volume of 2 $h^{-3}$Mpc$^{3}$. Smaller groups were excluded as voids, as they were not sufficiently resolved with our chosen cell size. 

\begin{figure*}
\includegraphics[trim={0cm 0cm 0cm 3.8cm}, clip, width=\linewidth]{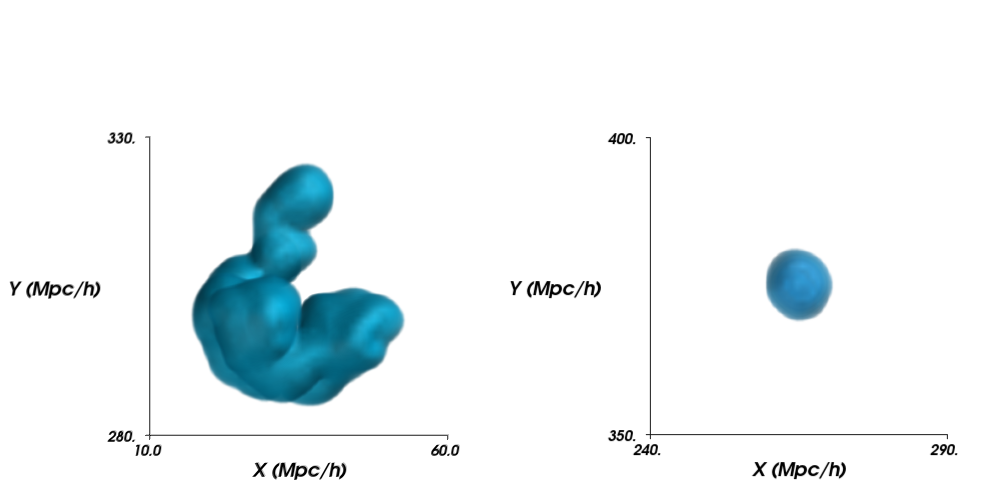}
\caption{An example of a void of volume 1899 $h^{-3}$Mpc$^{3}$ (left) and a void of volume 59 $h^{-3}$Mpc$^{3}$ (right) found in the $\Lambda$CDM simulation at $z=0.0$, using the Hessian-based method. Both voids are depicted using the 1 $h^{-1}$Mpc thick boundary sheet layer, hence they appear slightly larger than their volumes (which are calculated using void-like cells only) suggest. For clarity, only the $x$ and $y$ axes are shown. The $z$-axis points out of the page.} 
\label{Voldemort} 
\end{figure*}

It is worth noting that our Hessian-based method of void finding is different from other void finding methods in that it does not assume any particular shape, nor apply any density criterion, when identifying voids. There have been numerous methods used to define and identify voids. Some are based on the velocity field \citep[e.g. VWeb;][]{2012MNRAS.425.2049H}, but most methods use the density field. ZOBOV/VIDE \citep{2008MNRAS.386.2101N, 2014ascl.soft07014S} uses a watershed algorithm to define voids and requires that voids satisfy a minimum density criterion. Other density-based algorithms demand that voids be spherical underdensities \citep[e.g.][]{2005MNRAS.363..977P}. In contrast to these and several other commonly used methods, we do not enforce density thresholds when indentifying voids nor do we require voids to have any particular shape. As we do not apply any density constraints, we can identify voids that would be ignored by ZOBOV/VIDE. More importantly, not constraining the shape of voids enables us to find voids that are irregular in shape. We show an example of such a void in Figure \ref{Voldemort}, identified in the $\Lambda$CDM simulation at $z=0.0$. 

We find that the smallest voids (of a few $h^{-3}$Mpc$^{3}$ large) tend to be more regular (but not necessarily more spheroidal) in shape than larger voids. Since irregular shapes require more than a few cells to define, irregularity becomes more probable as voids get larger. The most likely explanation for this is void merging, with the largest voids forming as a result of the merger of multiple smaller and more regular voids. Figure \ref{Voldemort} shows an example of two voids found in the $\Lambda$CDM simulation using the method described above; one large, irregular void and one small, regular void. 

\section{Results}\label{Bellatrix} 
The three cosmologies were found to contain a very similar number of voids at $z=0.0$; a total of $35221$ voids were found in the $\Lambda$CDM simulation, compared $35701$ voids in the $\phi$CDM simulation and $34646$ voids in the CDE simulation. The similarity in numbers is not entirely surprising, since the initial conditions for each simulation contained the same initial density perturbation phases, and therefore the same initial underdensities. The differences in cosmology, which govern the initial amplitudes and the evolution of these underdensities, have resulted in a difference of only a few hundred voids. 

In this section, we will discuss the differences in the volume distribution, shape distribution and average density distribution of voids occurring in the three simulations across the low redshift universe, spanning $z=0.0$, $0.6$ and $1.0$. We will also discuss whether any of these differences could be used to observationally distinguish among the cosmologies, and hence differing dark sector physics. 
\subsection{Volume Distribution} 
A probability density function (PDF) to quantify the probability of finding a void of a specific volume was calculated for each simulation at multiple redshifts. We define the PDF so that the probability of finding a void of a given volume is the value of the PDF at that volume. We chose to use a PDF to quantify the void size distribution because it is independent of the total number of voids we found, and hence the size of the simulation box. The number of voids of a given volume within a region of space can be calculated by multiplying the PDF by the total number of voids we find (or expect to find) in that region of space, and reading off the value of the function at the given volume. 

We defined the volume of a void to be the number of 1 $h^{-3}$Mpc$^{3}$ void-like cells that comprise it, excluding the sheet-like cells that make up the boundary layer. To find the appropriate PDF, we plotted the number of voids found at each integer volume. The shape of this distribution showed a power-law dependence on volume for small voids, before falling more drastically with volume for larger voids, indicating that the underlying PDF has the same dependence. We do not attempt to follow the full excursion set formalism \citep[e.g.][]{2004MNRAS.350..517S}, but instead appeal to it for the rough functional form of our PDF, given by: 
\begin{equation}\label{powerlaw} 
f(V) = \frac{1}{V_{0} \Gamma(1-\alpha)}(V/V_{0})^{-\alpha} \mathrm{exp}(-V/V_{0}), 
\end{equation} 
where $V$ ($h^{-3}$Mpc$^{3}$) is the void volume, $\alpha$ is the power-law slope,  $V_{0}$ is the characteristic volume that defines the position of the turnover, and $\Gamma (1-\alpha)$ is simply the gamma function evaluated at $1-\alpha$: 
\begin{equation}
\Gamma(1-\alpha)=\int^{\infty}_{0} x^{-\alpha} e^{-x} dx.  
\end{equation}

\begin{figure}
\includegraphics[trim={0.5cm 0.5cm 0cm 0cm}, clip, width=\linewidth]{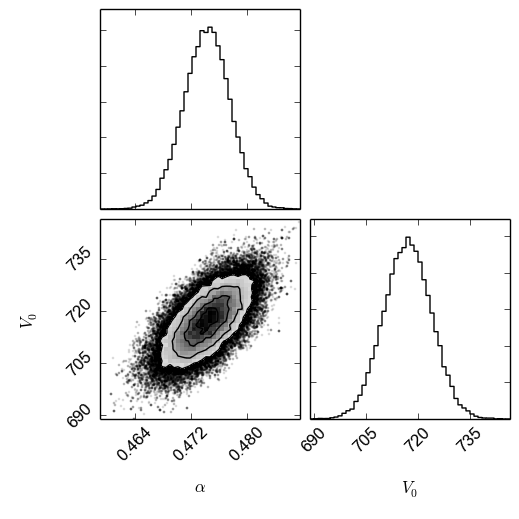}
\caption{Triangle plot showing the sampled parameter distribution for the $\Lambda$CDM volume PDF at $z=0.0$.}
\label{Katara} 
\end{figure}

\begin{figure*}
\includegraphics[width=\linewidth]{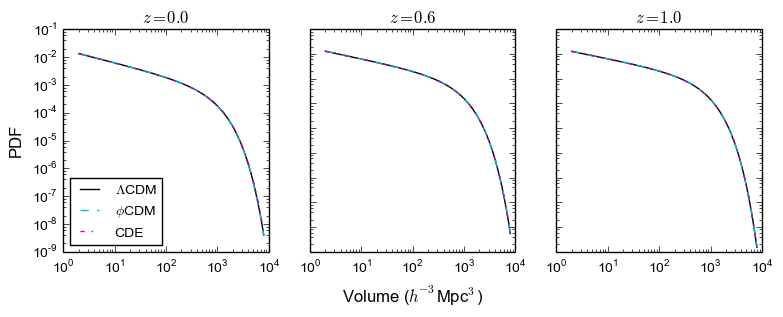} 
\caption{The void volume probability density functions at multiple redshifts for the $\Lambda$CDM (black solid), the $\phi$CDM (cyan dashed) and the CDE (magenta dot-dashed) simulations.}
\label{Harry}
\end{figure*}

Using a Metropolis-Hastings based Markov Chain Monte Carlo (MCMC) algorithm via the open source emcee package~\citep{emcee}, we obtained the best fit values and their uncertainties for the parameters $\alpha$ and $V_{0}$. These values were obtained for the volume PDFs at $z=0.0$, $z=0.6$ and $z=1.0$, for each of the three cosmological models, and are displayed in Table \ref{parameters}. The corresponding distributions are shown in Figure \ref{Harry}, while an example of a triangle plot obtained for the fit of the $\Lambda$CDM volume PDF at $z=0.0$ is presented in Figure~\ref{Katara}. 

\begin{figure*}
\includegraphics[width=\linewidth]{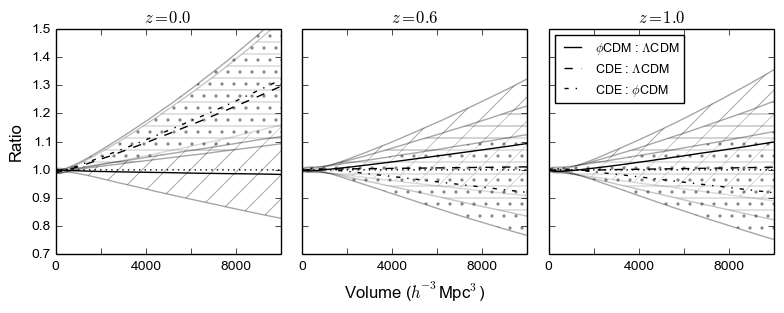}
\caption{The median ratios of the volume PDFs: $\phi$CDM/$\Lambda$CDM (solid line), CDE/$\Lambda$CDM (dashed line) and CDE/$\phi$CDM (dot-dashed line). The shaded regions show the range covered by the 16\textsuperscript{th} to 84\textsuperscript{th} percentiles of the ratio distributions, with the $\phi$CDM/$\Lambda$CDM region indicated by diagonal lines, the CDE/$\Lambda$CDM region indicated by horizontal lines, and the CDE/$\phi$CDM region indicated by dots. The black dotted horizontal line indicates a ratio of one.} 
\label{Aang} 
\end{figure*}

In each of the redshift snapshots, the PDFs for the three cosmologies display remarkable similarity over the volume range. The 1$\sigma$ uncertainties on their fit parameters reveal that the curves are consistent with each other over a large range of volumes. However, as Figure~\ref{Katara} shows, these fits are partially correlated, meaning that the best fit values and uncertainties obtained may not reflect the true parameter distribution. To determine if the PDFs are truly indistinguishable from each other, we performed a bootstrapping analysis on the distribution of values that were sampled during the fitting process. This involved taking a subsample of parameter values from the values that were sampled during the fitting, determining the corresponding PDFs from these subsamples, then calculating the ratios between PDFs of different models, thus producing a distribution of PDF ratios for each pair of models. We used 5000 subsamples to produce the PDF ratio distributions for each pair of models, which our convergence analysis showed would sufficiently represent the full ratio distributions. These distributions are represented in Figure~\ref{Aang}, which displays both the medians and the ranges covered by the 16\textsuperscript{th} to 84\textsuperscript{th} percentiles of the ratio distributions. The median ratios are displayed by the solid lines, while the interquartile ranges are represented by the shaded regions. As the ratio distributions are not Gaussian, we avoid using the 1$\sigma$ uncertainty, and instead use the 16\textsuperscript{th} to 84\textsuperscript{th} percentile range (which covers the same percentile range as 1$\sigma$) to represent the uncertainty in the median ratios. The increase in the spread of this range that we see towards high volumes may be attributed to small number counts for large voids, which means the shape of the PDF in this region is less constrained.  

The PDF ratios at $z=0.6$ and $z=1.0$ all tend to be consistent with a ratio of 1. The highest median ratio is $\approx1.1$ (between $\phi$CDM and $\Lambda$CDM), the lowest median ratio is $\approx0.93$ (between CDE and $\phi$CDM), and the corresponding interquartile ranges all safely overlap with a ratio of unity.  The ratio distributions between the CDE and the $\Lambda$CDM PDFs are the most consistent with 1, hence we would expect to see very little difference in void abundance across all volumes between these cosmological models. Though the medians do show greater deviation from unity at higher volumes between $\phi$CDM and $\Lambda$CDM, suggesting that on average, we should expect to see more voids with volumes of order $\sim10^{3}$ $h^{-3}$Mpc$^{3}$ in a $\phi$CDM than in a $\Lambda$CDM cosmology at these redshifts, the differences are unlikely to be distinguishable, especially given the considerable overlap with a ratio of unity in these distributions. The void abundance PDFs at these redshifts are therefore unlikely to be useful as probes of cosmology. 

However, there is a large range of volumes at $z=0.0$ where the CDE PDF deviates from the other two PDFs. The deviation from $\Lambda$CDM and $\phi$CDM starts at $V\approx700$ $h^{-3}$Mpc$^{3}$ and $V\approx880$ $h^{-3}$Mpc$^{3}$ respectively, and increases with volume. More than half of the ratios in the CDE:$\Lambda$CDM and the CDE:$\phi$CDM ratio distributions are greater than $1$. At volumes greater than $V\approx7\times10^{3}$ $h^{-3}$Mpc$^{3}$, more than half the ratios are greater than $1.1$. Additionally, we can reasonably expect to see at least $10\%$ more voids with volumes of $\approx4\times10^{3}$ $h^{-3}$Mpc$^{3}$ in a CDE universe than a $\Lambda$CDM or a $\phi$CDM universe, and this excess increases with volume. On the other hand, the distribution of ratios between the $\phi$CDM and $\Lambda$CDM PDFs show clearly that the void abundance in these cosmologies would be practically indistinguishable. 

Finally, the PDFs all evolve slowly with redshift. We note that $V_{0}$ increases with redshift for all cosmologies (with a significance of many $\sigma$), corresponding to an increase in the volume at which the turnover occurs. This is not entirely surprising, as we expect smaller voids to expand or merge to form larger voids, which would push the turnover out to larger volumes. The power-law index, $\alpha$, also slightly increases with redshift, which corresponds to a slight increase in the steepness of the power-law. However, as both parameters are correlated (Figure~\ref{Katara}), the change we observe in $\alpha$ may simply reflect the change in $V_{0}$ from $z=1.0$ to $z=0.0$. The similarity in power-law slope across redshifts is consistent with excursion set-based void abundance predictions that do not show a large power-law dependence on redshift~\citep{2013MNRAS.434.2167J}. 

\begin{table*}
\caption{The best fit values (including 1$\sigma$ errors) for the volume probability density function parameters, $\alpha$ and $V_{0}$. The values for three redshifts are shown for each cosmology.} 
\begin{center}
\begin{tabular}{ c|c|c|c|c }
Cosmology & Params & $\textbf{z=0.0}$ & $\textbf{z=0.6}$ & $\textbf{z=1.0}$ \\
\hline 
\hline
 $\Lambda$CDM &  $\alpha$ & $0.474\pm0.003$ & $0.442\pm0.003$ & $0.423\pm0.003$ \\
& & & &\\
 & $V_{0}$ & $717.1^{+6.8}_{-6.9}$ & $598.3^{+5.4}_{-5.3}$ & $540.0^{+4.7}_{-4.7}$\\
\hline 
 $\phi$CDM &  $\alpha$ & $0.476\pm0.003$ & $0.445\pm0.003$ & $0.427\pm0.003$ \\
& & & &\\
 & $V_{0}$ & $716.2^{+6.9}_{-6.9}$ & $602.8^{+5.5}_{-5.4}$ & $543.2^{+4.8}_{-4.7}$\\
\hline
 CDE & $\alpha$ & $0.477\pm0.003$ & $0.441\pm0.003$ & $0.424\pm0.003$\\ 
& & & &\\
 & $V_{0}$ & $731.7^{+7.1}_{-7.1}$ & $599.0^{+5.5}_{-5.3}$& $540.6^{+4.7}_{-4.6}$\\
\hline
\end{tabular}
\end{center}
\label{parameters} 
\end{table*}

\subsection{Shapes} 
Voids are often assumed to be roughly spherical, for example during the void finding process \citep[as in][]{2005MNRAS.363..977P}, and in the construction of density profiles as a function of radius from the void centre \citep{2014PhRvL.112y1302H}. However, our simulations suggest that voids can be quite irregular in shape (for example, Figure \ref{Voldemort}), and are better approximated by ellipsoids than spheroids, especially for the smaller, unmerged voids. We define the ellipticity of a void (or deviation from a perfect sphere), $e$, to be: 
\begin{equation}\label{Dumbledore} 
e = \frac{1}{4} \frac{c^{2} - a^{2}}{a^{2} + b^{2} + c^{2}},  
\end{equation}
where $a$, $b$ and $c$ are the three axes that define the ellipsoid, with $c$ being the longest and $a$ being the shortest. This definition is identical to the definition given by \citet{2016MNRAS.461..358N}, except that we express $e$ in terms of the three ellipsoid axes, rather than eigenvalues of the inertia tensor. With this definition, a perfect sphere has an ellipticity of zero. The greater the value of $e$, the greater the deviation from a perfect sphere. 

Ellipsoids can be either prolate (more elongated along one axis) or oblate (more elongated along two axes). To quantify this variation in shape, we define the prolateness, $p$, to be: 
\begin{equation}\label{prolate}
p = \frac{1}{4} \frac{(b^{2} - a^{2}) + (b^{2} - c^{2})}{a^{2} + b^{2} + c^{2}}. 
\end{equation} 
Again, this definition is identical to that given by \citet{2016MNRAS.461..358N}. A negative value of $p$ indicates elongation along one axis relative to the other two (prolate), whereas a positive value of $p$ indicates relative elongation along two axes compared to the third axis (oblate). A prolateness of zero indicates that the ratio between the shortest axes and the longest axes are comparable, meaning that the shape is neither prolate nor oblate. 

\begin{figure*} 
\includegraphics[width=\linewidth]{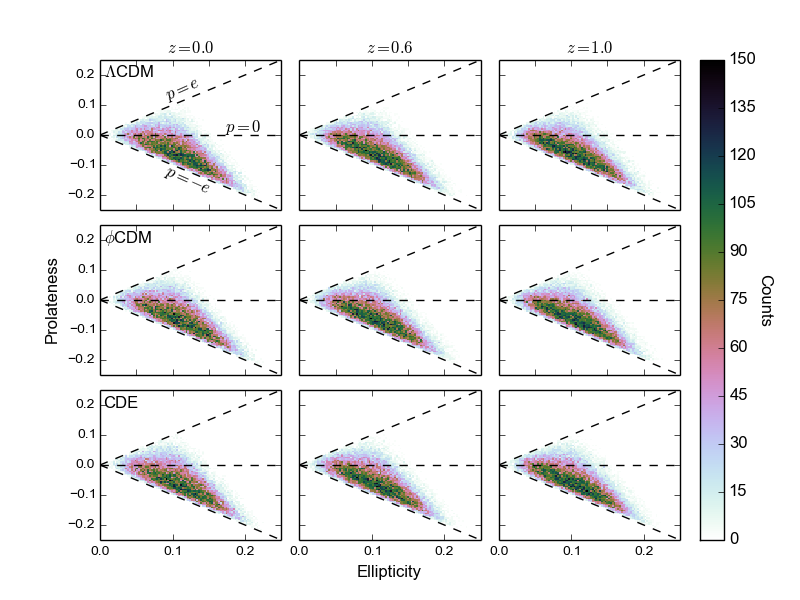}
\caption{Distribution of void ellipticity versus prolateness in $\Lambda$CDM (top row), $\phi$CDM (middle row) and CDE (bottom row), for $z=0.0$, $z=0.6$ and $z=1.0$. The horizontal black dashed line represents a prolateness of zero. The $p=e$ line represents the limiting case where the two shortest sides are equal in length (i.e. $a=b$), while the $p=-e$ line denotes the limiting case where the two longest sides are equal in length (i.e. $b=c$). }  
\label{Dobby} 
\end{figure*}

We calculated the ellipticities and prolateness of the voids in each of the three simulations, by calculating the eigenvalues of the moment tensor for each void, given by 
\begin{equation}
M_{ab} = \sum_{i} (x_{i}^{a} - X^{a})(x_{i}^{b} - X^{b}), 
\end{equation} 
where $i$ denotes the sheet-like cells that comprise the boundary layer of the void, $x_{i}^{a}$ and $x_{i}^{b}$ denote the $a$-coordinate and $b$-coordinate of the $i$-th boundary layer cell respectively, and $X^{a}$ and $X^{b}$ are the $a$-coordinate and $b$-coordinate of the barycentre of the void respectively, with both $a$ and $b$ representing the $x$, $y$ and $z$ axes. The eigenvalues of this tensor ($e_{1}$, $e_{2}$, $e_{3}$) correspond to a third of the square of the axes of the ellipsoid that best approximates the shape of the void: 
\begin{equation}
e_{1} = \frac{a^{2}}{3}, ~e_{2} = \frac{b^{2}}{3}, ~e_{3} = \frac{c^{2}}{3}.  
\end{equation} 
From these eigenvalues, we calculated the length of the ellipsoid axes, and thus the ellipiticities and the prolateness of each void. We note that we included only voids of size greater than or equal to 10 cells (equivalent to 10 $h^{-3}$Mpc$^{3}$ in volume) in this analysis, since voids smaller than this do not have resolved shapes. Figure \ref{Dobby} shows the resulting distributions of prolateness versus ellipticity, for the voids in $\Lambda$CDM, $\phi$CDM and CDE at $z=0.0$, $0.6$ and $1.0$. The distributions are all peaked at ($e$, $p$) = (0.10, -0.05), with 25\textsuperscript{th} and 75\textsuperscript{th} percentiles occuring at ($e$, $p$) = (0.08, -0.09) and ($e$, $p$)  = (0.13, -0.02) respectively, consistent with the results derived in \citet{2016MNRAS.461..358N} for voids in the SDSS-III BOSS Data Release 11 \citep{2015ApJS..219...12A}. There is no significant difference among the cosmologies in the distribution of ellipticities or prolateness, with the most common void shape across all three cosmologies being slightly prolate. 

\begin{figure}
\includegraphics[trim={1cm 0cm 1cm 1cm},clip, width=\linewidth]{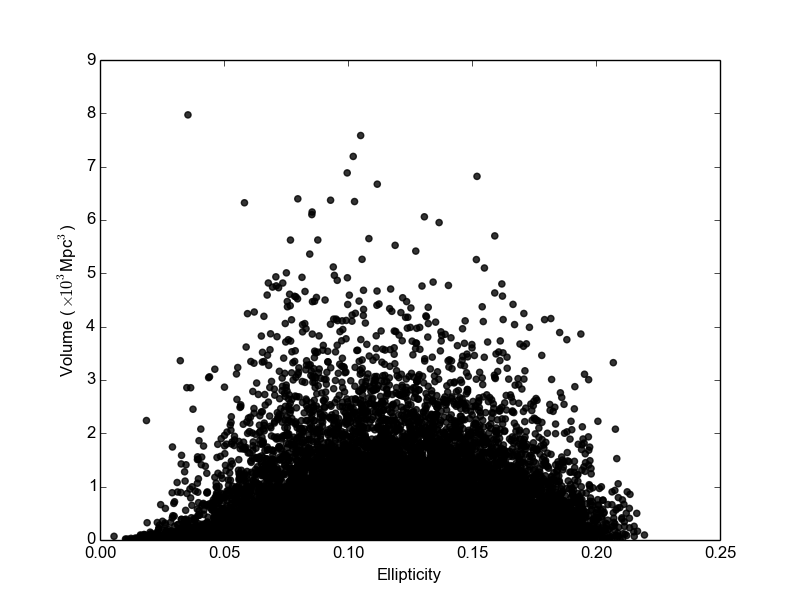}
\caption{Relationship between void size and ellipticity for $\Lambda$CDM at $z=0$. The distribution shows that the smallest voids tend to have a larger spread of ellipticities than the larger voids.}
\label{Ginevra}
\end{figure}

As mentioned in Section~\ref{Luna}, the largest voids can be very irregular due to their formation through the merger of smaller voids. Interestingly, the irregularity of large voids does not automatically translate to a noticeable deviation from an approximate spheroid, which is clear from the distribution of void size versus ellipticity in Figure~\ref{Ginevra}. In fact, the smaller voids tend to exhibit a greater spread in ellipticity than the larger voids. This result is consistent with previous studies into the spherical evolution of voids, which found that the largest voids are better approximated by a spherical model than the smallest voids \citep[e.g.][]{2015MNRAS.451.3964A, 2016MNRAS.463..512D}. A likely reason for this is that there are many more ways to merge smaller voids to form a lower ellipticity void than a high ellipticity void, and hence a greater number of approximately spheroidal large voids form than approximately ellipsoidal large voids. 


\subsection{Densities} 
\begin{figure*}
\includegraphics[width=\linewidth]{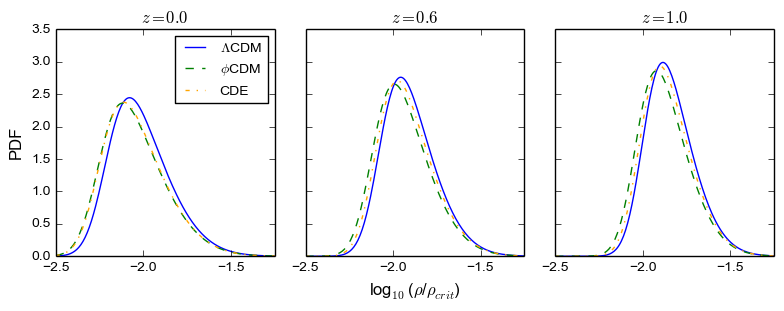}
\caption{Probability distribution functions for the base 10 logarithm of the average void density in the $\Lambda$CDM (solid blue), $\phi$CDM (dashed green) and CDE (orange dot-dashed) cosmological simulations at $z=0.0$, $z=0.6$ and $z=1.0$, in units of the critical density. The $\Lambda$CDM spherical threshold corresponds to a vertical line at log$_{10}$($\rho/\rho_\mathrm{{crit}}) = -1.27$.} 
\label{Hermione} 
\end{figure*}

For each simulation and at multiple redshifts, we calculated a PDF to quantify the probability of finding a void of a specific average density. The average density for each void was calculated by taking the average of the smoothed densities of each cell in a void (excluding the boundary layer of sheet-like cells), which we calculated during the cell classification process. We found that the best fitting function for the underlying PDF governing the distribution of average void densities was a skewed Gaussian distribution, defined as a function of the base 10 logarithm of the void density, given by 
\begin{equation}
p(t)=\frac{1}{\sqrt{2\pi}}e^{-t^{2}/2}\left[1 + \mathrm{erf} \left(\frac{\alpha t}{\sqrt{2}}\right)\right], 
\end{equation}
where $\alpha$ is the skewness parameter and $t\equiv(x - \mu)/\sigma$, with $\mu$ and $\sigma$ respectively denoting the mean and the standard deviation of the corresponding Gaussian. Again we used a Metropolis-Hastings based MCMC fitting algorithm to obtain the best fit values for these parameters, which are summarised in Table \ref{densityparameters}. Figure~\ref{Hermione} shows the shapes of the PDFs with these parameter values, highlighting the differences in the average void density distribution among the three cosmological simulations at $z=0.0$, $z=0.6$ and $z=1.0$.  The spherical threshold, which denotes the maximum density defined for voids predicted by excursion set theory, is given by $\Delta_{v}=-0.8$ for $\Lambda$CDM, or equivalently log$_{10} (\rho/\rho_{\mathrm{crit}}) = -1.27$ in the units chosen for Figure~\ref{Hermione}. It is clear from the figures that the vast majority of our voids in all three simulations have average densities below this threshold. 

\begin{figure*}
\includegraphics[width=\linewidth]{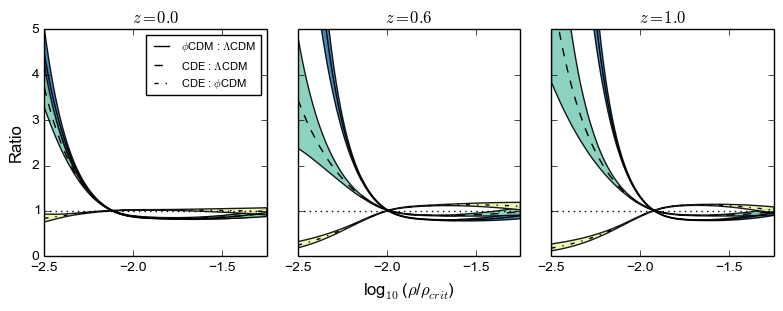} 
\caption{The median ratios of the density PDFs: $\phi$CDM/$\Lambda$CDM (solid), CDE/$\Lambda$CDM (dashed) and CDE/$\phi$CDM (dot-dashed). The shaded regions show the range covered by the 16\textsuperscript{th} to 84\textsuperscript{th} percentiles of the ratio distributions, with $\phi$CDM/$\Lambda$CDM in dark blue (or dark grey in black-and-white print), CDE/$\Lambda$CDM in green (or medium grey) and CDE/$\phi$CDM in yellow (or light grey). A ratio of unity is displayed by the black dotted horizontal line.} 
\label{Granger}
\end{figure*}

To show the differences among the simulations more rigorously, we once again subsampled the parameter distributions from the fitting process, and used the subsamples to calculate corresponding distributions of PDF ratios. As before, our convergence analysis showed that 5000 subsamples sufficiently represents the full ratio distributions. Figure~\ref{Granger} shows the median of these PDF ratios, along with the 16\textsuperscript{th} and 84\textsuperscript{th} percentiles, for $z=0.0$, $0.6$ and $1.0$. 
 
It is clear from Figures~\ref{Hermione} and~\ref{Granger} that the PDFs are almost all distinct from one another for all three redshifts. At $z=0.6$ and $z=1.0$, the $\Lambda$CDM PDF has the highest density peak, while the $\phi$CDM PDF has the lowest. This indicates that on average, the voids in both $\phi$CDM and CDE cosmologies are slightly emptier than the voids in a $\Lambda$CDM cosmology. This is also seen at $z=0.0$, where the $\phi$CDM and CDE PDFs appear to have converged, while $\Lambda$CDM remains distinct. The PDF peaks (characterised by $\mu$) also shift towards lower densities as the simulations evolve from $z=1.0$ to $z=0.0$, while their spread (given by $\sigma$) and skewness towards the lower density region (given by $\alpha$) generally increases. This suggests that voids on average become emptier with time, as expected, since matter continuously evacuates out of voids as it accumulates onto higher density structures. The deviations from a ratio of 1 at low densities is due to the ratios of small numbers, since the corresponding PDF values drop off to zero at slightly different rates. The exact shape of drop-offs are unlikely to be well-characterised though, due to small number statistics in this region. For this reason, the deviations at the lowest densities are unlikely to be useful in distinguishing among cosmological models.  

\begin{table*} 
\caption{The best fit values (including 1$\sigma$ errors) for the skewed Gaussian parameters, $\mu$, $\sigma$ and $\alpha$. The values for three redshifts are shown for each cosmology.} 
\begin{center}
\begin{tabular}{ c|c|c|c|c }
Cosmology & Params & $\textbf{z=0.0}$ & $\textbf{z=0.6}$ & $\textbf{z=1.0}$ \\
\hline 
\hline
 $\Lambda$CDM & $\mu$ & $-2.208\pm0.001$ &  $-2.071\pm0.001$& $-1.991\pm0.001$\\  
& & & &\\
 & $\sigma$ & $0.260\pm0.001$ & $0.226\pm0.001$ & $0.205\pm0.001$ \\
& & & &\\
 & $\alpha$ & $2.600^{+0.044}_{-0.001}$ & $2.415^{+0.039}_{-0.001}$ & $2.256^{+0.036}_{-0.001}$\\
\hline 
 $\phi$CDM & $\mu$ & $-2.251\pm0.001$&  $-2.113\pm0.001$& $-2.033\pm0.001$\\  
& & & &\\
 & $\sigma$ & $0.272\pm0.001$ & $0.237\pm0.001$ & $0.217\pm0.001$ \\
& & & &\\
 & $\alpha$ & $2.715^{+0.046}_{-0.001}$ & $2.493^{+0.041}_{-0.001}$ & $2.368^{+0.037}_{-0.001}$\\
\hline
 CDE & $\mu$ & $-2.246\pm0.001$ & $-2.092\pm0.001$ & $-2.010\pm0.001$\\  
& & & &\\
 & $\sigma$ & $0.270\pm0.001$ & $0.232\pm0.001$ & $0.210\pm0.001$\\ 
& & & &\\
 & $\alpha$ & $2.700^{+0.047}_{-0.001}$ & $2.474^{+0.041}_{-0.001}$& $2.295^{+0.036}_{-0.001}$\\
\hline
\end{tabular}
\end{center}
\label{densityparameters} 
\end{table*} 

\begin{figure}
\includegraphics[trim={0.3cm 0.4cm 0cm 0cm}, clip, width=\linewidth]{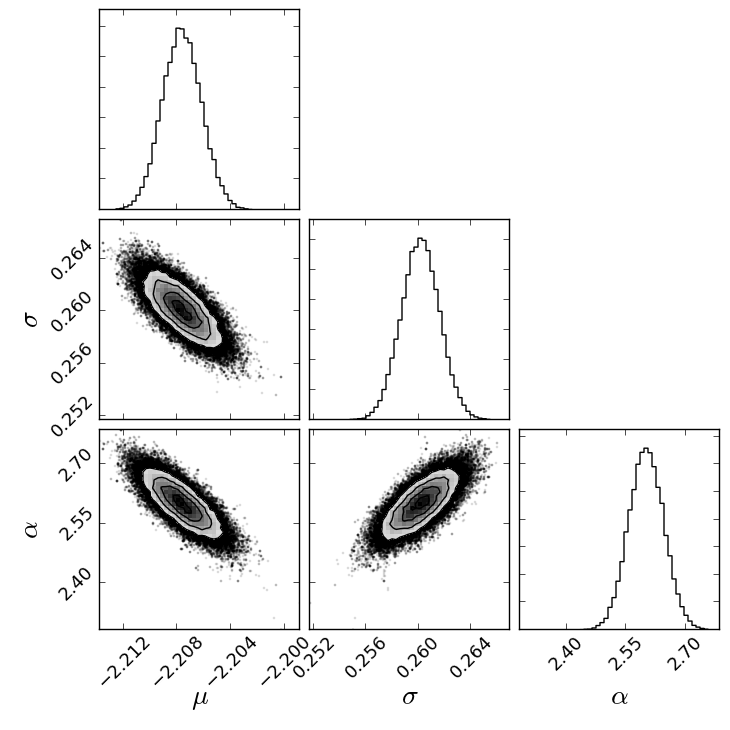}
\caption{Triangle plot showing the sampled parameter distribution for the $\Lambda$CDM density PDF at $z=0.0$.}
\label{Zuko} 
\end{figure} 

\section{Discussion} 
Our simulations show that changes in expansion history and the presence of coupling in the dark sector alters the void size distribution at $z=0.0$, as well as the average density of voids across $z=0.0-1.0$. However, these differences in dark sector physics do not alter the void ellipticity or prolateness distribution at any redshift, nor does it affect the void size distribution at $z=0.6-1.0$. Given that our simulations differ from each other primarily in the effects of the underlying cosmology, the differences we see among void properties are significant and can be confidently attributed to differences in the underlying model (and hence dark sector physics). We may view these differences as imprints or signatures of dark sector physics on the cosmic web, and use them to observationally constrain the properties of the dark sector. The most promising signature we have found is the potential for the average void density distributions to distinguish between the standard cosmological model and the evolving dark sector models. 

\subsection{Volume Distribution} 
The median ratio between $\phi$CDM and $\Lambda$CDM is much more consistent with unity at $z=0.0$ than at the higher redshifts, potentially signalling greater discrepancies with increasing redshift. At $z=0.6-1.0$, the ratio reaches $\approx1.05$ at volumes of around $\approx5\times10^{3}$ $h^{-3}$Mpc$^{3}$, and continues to rise with volume. Though these ratios only show a small deviation from unity at low redshifts, they hint at the existence of greater discrepancies between $\phi$CDM and $\Lambda$CDM at redshifts greater than $z=1.0$. However, to determine whether this is the case, the same analysis must be conducted on simulation snapshots at a range of redshifts greater than $z=1.0$, which will be covered in a follow-up study (Adermann et al., in prep.). 
%

The lack of significant differences between the void volume distributions of $\phi$CDM and $\Lambda$CDM suggests that for our particular form of $\phi$CDM, the scalar field does not leave a strong imprint on the abundance of voids in the universe at late times. The specific form of $\phi$CDM we chose exhibits only small changes in the expansion history compared to $\Lambda$CDM, with the greatest deviation corresponding to a Hubble parameter approximately 1.02 times greater than in the $\Lambda$CDM model, occuring at late times (see Figure~\ref{Sims}). This deviation may have been too small to affect the volume distribution significantly, or the change may have occurred too late in the formation of each void for their sizes to be significantly influenced. This is especially important to consider if final void sizes are most heavily influenced by conditions early in the process of void formation. For our $\phi$CDM model, the final void sizes appear to be dominated by the initial density perturbations and gravity, and only minimally influenced by expansion history, or equivalently, the scalar field. As there are multiple realisations of a $\phi$CDM universe, these results do not rule out more visible signatures on void abundance under $\phi$CDM models with different forms of the potential $V(\phi)$. 

Our results also show that the CDE model leaves an imprint on the void size distribution at $z=0.0$, while $\phi$CDM is indistinguishable from $\Lambda$CDM at the same redshift. Therefore, it is the coupling between the scalar field and dark matter that most affects the abundance of voids in the universe today, rather than the presence of the scalar field in place of a cosmological constant. For our specific models, the coupling results in an increase in the abundance of voids with $V\sim10^{3}$ $h^{-3}$Mpc$^{3}$. Interestingly, the effect of the coupling is only apparent at $z=0.0$. At higher redshifts, its effect is much less noticeable. 

These results are partially consistent with the results obtained by \citet{2015MNRAS.446L...1S}, who found an increased number of large voids in their coupled dark sector model compared to $\Lambda$CDM, at $z=0$. However, the excess they show in void abundance occurs at larger volumes than indicated by our results. While our results show a greater void abundance for voids of order $\sim10^{3}$ $h^{-3}$Mpc$^{3}$, their results show a consistency in the abundance of voids this size between $\Lambda$CDM and CDE, with the excess starting at effective radii of $>25-30$ $h^{-1}$Mpc, corresponding to volumes of $\sim10^{4}$ $h^{-3}$Mpc$^{3}$. It is worth noting that the coupled model used in this study is governed by the same Lagrangian as ours, apart from the use of a greater coupling constant \cite[cDE with $\beta=0.099$; for model specifics, see][]{2015MNRAS.446L...1S}.  

\citet{2015MNRAS.446L...1S} suggest that the reason for the excess is the thinning of void walls when moving from the standard to the coupled dark sector models. Smaller voids in the $\Lambda$CDM cosmology may then be identified as a single larger, merged void in the coupled dark sector cosmology. In our early analyses of the differences in voids between our $\Lambda$CDM and $\phi$CDM models (at $z=0.0$) in 50 $h^{-1}$Mpc and $2\times128^{3}$ particle simulations, we discovered that some of the largest voids in the $\phi$CDM model correspond to two or three smaller voids in the $\Lambda$CDM model, that had been merged in the $\phi$CDM model, consistent with \citet{2015MNRAS.446L...1S}'s suggestion. This suggests that thinning of void walls may be linked to the scalar field, which is present in both the $\phi$CDM and CDE models, in addition to its coupling to dark matter. 

However, this effect is not significant in our current results, when we applied the same analyses to the larger 500 $h^{-1}$Mpc and $2\times 1024^{3}$ particle simulations. Despite the evidence for wall thinning and merging in the smaller simulations, the larger $\phi$CDM simulation does not predict a significant excess in large voids compared to $\Lambda$CDM at $z=0.0$, which suggests that though it may occur, wall thinning may not be drastic enough to cause most neighbouring voids to merge. On the other hand, the relative excess in larger voids between CDE and $\Lambda$CDM is consistent with more significant wall thinning occurring between voids, as found by \citet{2015MNRAS.446L...1S}. This suggests that the coupling to dark matter by the scalar field has a much more noticeable effect on wall thinning than the scalar field on its own. This might be expected as the additional effective drag acting between baryonic matter and dark matter due to the coupling would potentially slow down the evacuating matter, leaving less material to accumulate onto the void walls. 

Despite the similarities in cosmological models used by \citet{2015MNRAS.446L...1S}, there is an interesting difference between their results and ours; namely, the volume range at which CDE produces an excess of voids relative to $\Lambda$CDM. We find the excess to be statistically significant but not especially large (up to $\approx1.3$ times as many as $\Lambda$CDM), whereas \citet{2015MNRAS.446L...1S} find several times as many voids in their CDE model than in their $\Lambda$CDM model. There are a number of reasons for this. Firstly, \citet{2015MNRAS.446L...1S} use a stronger coupling strength ($\beta=0.099$ compared to our value of $\beta=0.05$). It seems likely that the excess (both amplitude and range) depends on the coupling, though we cannot say for certain without a larger parameter space studied. Increasing coupling leads to an increased drag force, which in turn results in thinner void walls. Thinner walls increase void merger rates and artificial merging of voids by void finders, thereby creating a greater discrepancy between CDE and $\Lambda$CDM. Without detailed tracking of void evolution as was done in \citet{2014MNRAS.445.1235S}, this explanation remains untested. In future studies, we will follow the methodology of \citet{2014MNRAS.445.1235S} and produce void merger trees akin to halo merger trees. 

There are also a few potential reasons for the differences between their results and ours that do not imply a physical origin: the simulation volumes differ (theirs was an eighth the volume of ours, with higher mass resolution); and they used a watershed-based algorithm (VIDE) to identify voids in their simulations (biases in void finding techniques remain to be thoroughly catalogued). Although differences in mass resolution result in differences in the density field, at large-scales the low density regions that correspond to voids are generally well-converged and independent of resolution. Small scales, and hence small voids, are more affected by resolution, but they are unlikely to be a major cause of the differences because we are neglecting the smallest, unresolved voids from our analyses. Differences could also arise due to the use of other criteria in their void finding that we did not impose, such as a density and size criterion, as well as the use of different smoothing scales to calculate densities. However, these should not alter the void finding results significantly, because imposing other criteria should only affect a small population of the densest and smallest voids, while using different smoothing scales will only affect voids that are in the process of merging.  Thus, we believe that the physical origin, namely the coupling strengths, remains the most reasonable explanation and primary reason for the difference.

Our results are generally consistent with the results obtained by \citet{2016MNRAS.455.3075P}, who also found an excess of large cosmic voids in their coupled dark sector model compared to $\Lambda$CDM, over the redshift range $z=0.00-1.00$ \cite[for more details on their coupled model, see][] {2016MNRAS.455.3075P}. As with our results, this excess increases with void size. However, the excess was found to steadily increase with decreasing redshift, whereas our results show a more sudden increase at $z=0.0$. Furthermore, the distributions they derive do not decrease with increasing size, but rather peak at effective radii of roughly $20-30$ $h^{-1}$Mpc. The range of void sizes they find include voids with effective radii up to $\approx40$ $h^{-1}$Mpc in size, while the largest voids in our simulations were only $\approx10^{4}$ $h^{-3}$Mpc$^{3}$. 

The use of different CDE models may account for the discrepancies we observe; \citet{2016MNRAS.455.3075P} apply an exponential form for the potential governing the scalar field ($V(\phi)=A$exp$(-\alpha\phi)$), as well as a much greater coupling strength ($\beta=0.15$). Greater coupling, in addition to a different expansion history, may lead to a stronger signature at higher redshifts than $z=0.0$, accounting for the differences in when the signature starts showing up. The inconsistency in void sizes is likely to be the result of using different void finding methods. \citet{2016MNRAS.455.3075P} use the same void finder as \citet{2015MNRAS.446L...1S}, who also find voids with much larger sizes than we find. The watershed-based algorithm appears to identify much larger voids than our density-based method. In addition, \citet{2016MNRAS.455.3075P} apply additional cuts based on density contrast, which we do not apply. It is possible that many of their smallest voids were eliminated in this way, that were not eliminated in either \citet{2015MNRAS.446L...1S}'s void catalogue nor ours. 

The normalisation of the amplitudes of the density perturbations used in their simulations also differ from ours \citep[and that of][]{2015MNRAS.446L...1S}. While the density perturbations in the models we use are normalised to have the same $\sigma_{8}$ at $z=0.0$, those used in \citet{2016MNRAS.455.3075P} are normalised to have the same amplitude of perturbations at $z=z_{CMB}$, and consequently the amplitudes in their models differ most at $z=0$ \citep[for an in-depth discussion on normalisation method, see][]{2015MNRAS.452.1341E}. As we expect the amplitudes of density perturbations to affect the growth of voids, it is reasonable to expect that some of the deviations between CDE and $\Lambda$CDM seen in \citet{2016MNRAS.455.3075P}, including the fact that the deviation increases with decreasing redshift, could be due to the way the simulations were normalised. The best way to test this would be to derive the void volume distribution from models that have been normalised differently.

\subsection{Shapes} 
The remarkable similarity in the $e$ vs $p$ distributions across simulations indicates that changes in expansion history and dark matter physics do not drastically alter the shape of the initial underdensities, which are the same across all simulations. This result suggests that the overall shape of voids is barely affected by small changes to expansion history due to a scalar field or additional drag terms due to coupling within the dark sector but rather, is dominated by the effect of gravity and the initial phases of the density perturbations. However, we note that higher order shape statistics may carry more noticeable imprints, which are not captured in our shape statistics. 

Our results also indicate that the common assumption of sphericity for voids is problematic. In each of the shape distributions, there is a spread of $\approx0.2$ in the range of ellipticity and prolateness values, with a peak in the distributions at $e\approx0.10$ and $p\approx-0.05$. This demonstrates that voids, when approximated by a regular ellipsoid, match an ellipsoid shape more commonly than a spheroid shape ($e = 0$, $p = 0$). Though the general deviation from a spheroid is not huge, the commonly used spheroid assumption for stacked voids still results in a loss of information that may have implications for experiments involving the Sachs-Wolfe effect or the Alcock-Paczynski test. In particular, the spheroid assumption would be problematic if the overall ellipsoid shapes are not randomly oriented along different lines of sight, such that the deviations average out. We suggest that a more accurate method of stacking should assume that voids are ellipsoidal rather than spheroidal, drawn from a distribution in $e$ and $p$.  

The ellipsoid average has implications for defining and analysing void profiles. Void profiles are conventionally determined using an effective void radius, which implicitly assumes that the void can be approximated by a sphere. If most voids are ellipsoidal, then it may be more informative to define effective ellipsoid axes in which to calculate void density profiles, rather than use an effective radius, where density variation information is invariably lost. However, it should be noted that a void density profile will still be a somewhat crude measure of how density varies within highly irregular voids (which many of our voids are), even if the profiles are calculated with respect to effective ellipsoid axes. Profiles for such voids would ideally take the shape of the void into account. 

\subsection{Densities} 
Out of the three void properties we have explored thus far, the average void density distribution shows the most potential for use as a probe of cosmology.  Except for the similarity between $\phi$CDM and CDE at $z=0.0$, these distributions are all distinct from one another across redshifts $z=0.0-1.0$. In particular, at all three redshifts, $\Lambda$CDM  produces denser voids than the other cosmologies. 

A possible cause of the distinct difference in average void densities is the way dark energy is characterised in both $\phi$CDM and CDE compared to $\Lambda$CDM. In both the alternative models, the accelerated expansion of the universe arises due to the presence of a scalar field, whereas in the standard model, the cause is attributed to the presence of dark energy that produces an outwards pressure on the universe. The inclusion of an additional scalar field affects the expansion history of the universe, which in turn may affect the speed at which matter evacuates out of voids. Under the Ratra-Peebles potential used in our quintessence and coupled models, matter appears to evacuate out of voids more quickly.

The differences present in the average void densities can also be understood in terms of the non-linear power spectra in Figure~\ref{Sims}. Both $\phi$CDM and CDE have more power at scales where the vast majority of our voids are found ($0.1 \lesssim k \lesssim 1$) than $\Lambda$CDM. This is consistent with denser void walls and thus emptier voids, found in both the alternative models compared to the standard model. 

The $\phi$CDM and CDE PDFs are distinct from each other at $z=0.6$ and $z=1.0$, indicating that the presence of coupling does affect the speed of evacuation at some time during void evolution. Specifically, as the peak of the CDE PDF occurs at a higher density than the peak of the $\phi$CDM PDF at $z=0.6$ and $z=1.0$, the presence of coupling between the scalar field and dark matter appears to cause voids to evacuate more slowly and contain more matter. This is consistent with the additional drag force between particles, which may be expected to slow down the evacuation of matter from voids. However, despite the additional drag force, the CDE PDF still peaks at a lower density than the $\Lambda$CDM PDF, indicating that voids in a CDE cosmology are still on average emptier than the voids in a $\Lambda$CDM cosmology. One possible explanation for this is that the effect of the scalar field (which is also present in our CDE model) in emptying out voids is greater than the effect of coupling in slowing down void evacuation. The lower density peak is offset by the effect of coupling, but not enough to produce voids as dense as those in $\Lambda$CDM on average. 
 
Interestingly, the difference in the distributions reduces to insignificance at $z=0.0$, as the two PDFs converge. There are two possible explanations for this. The first is that the effect of coupling naturally reduces with time, perhaps as the evacuation speed in the CDE cosmology speeds up as fewer particles are left in voids, causing the average void densities in these two models to converge, and ultimately leaving no evidence of coupling on the average emptiness of voids in the present day universe. If true, this would be a remarkable coincidence. The second explanation is that the evolution of the void densities are influenced by the normalisation of the density field perturbation amplitudes (our simulations are all normalised so that the mass variance, $\sigma_{8}$ at $z=0.0$, is the same for all of them). Indeed, this would not be surprising given that the normalisation affects the evolution of the initial density field perturbations (i.e. if we chose a different redshift for the normalisation, then the evolution of those amplitudes would look different). If this effect is as significant as the effect of coupling between dark matter and the scalar field, then it is possible that the differences we see between the $\phi$CDM and CDE PDFs (and between the $\Lambda$CDM and CDE PDFs) across all three of these redshifts are affected by this normalisation. Determining the significance of the time of normalisation would require extracting the average void density PDFs from simulations normalised at different epochs, and checking for differences in the evolution of the PDFs. 

Fortunately, the differences between the $\Lambda$CDM PDF and the $\phi$CDM PDF are unlikely to be affected by our normalisation, since the presence of the scalar field alters only the expansion history, and not the evolution of the initial density perturbations. We can therefore conclude that the scalar field, or the nature of dark energy, imprints itself onto the average density of voids. In our specific models, it produces emptier voids in the universe. 

It is worth noting that the specific differences we see in the density PDFs may be very model-dependent. For example, using a different potential for the Lagrangians underpinning the $\phi$CDM and the CDE models may result in different PDF profiles, compared to $\Lambda$CDM. Furthermore, the choice of parameter values such as coupling strength may affect the size of the differences between the PDFs. For this reason, it is possible for certain realisations of alternative models to produce density PDFs that are very similar to that of the standard $\Lambda$CDM. Conversely, any difference found between the predicted $\Lambda$CDM PDF and the PDF derived from actual observations will not necessarily inform us of the specific cosmology of the universe. In order for strong constraints on the cosmology to be made, more studies into the particular dependence of the void densities on the specific model and its parameter values must be done. 

Once we have predictions from a wide range of models, the advantage of using the average void density PDF as a probe of cosmology is clear. Unlike the volume distributions, the differences between models in the average void density distributions occurs across the redshift range $z=0-1$, which corresponds to a significantly larger observational volume in which to search for signatures of cosmology.  

\section{Conclusions}
We have studied the properties of voids in three different cosmological models: the standard $\Lambda$CDM model, a quintessence model with a time-varying scalar field and a coupled dark energy model featuring coupling between a time-varying scalar field and dark matter. We have made comparisons of void properties between these models in the redshift range $z=0.0-1.0$, with the goal of identifying the signatures left on voids by the underlying dark sector physics within these models.  

We ran numerical $N$-body simulations of each of the three cosmological models in boxes of length $500$ $h^{-1}$Mpc, consisting of $512^{3}$ dark matter and baryonic particles each. Each simulation was initialised to have the same density perturbation phases, enabling voids between simulations/models to be matched and compared. The primary difference between the simulations lay in the governing cosmological model, which allowed direct comparisons of various void properties to be made among the models and any differences found to be attributed to differences in underlying cosmology. Using a Hessian-based void-finding algorithm without applying shape or density requirements, we were able to find a variety of voids in the cold dark matter distributions of these simulations, many of which are highly irregular and would not have been found via other methods. We conducted analyses on the sizes, shapes and average densities of these voids. For each of the three cosmological models, we derived a probability density function describing the void size distribution, a void ellipticity and prolateness distribution, and a probability density function describing the average void density distribution. 

We found that the CDE model is distinct from both $\Lambda$CDM and $\phi$CDM in the distribution of void sizes at $z=0.0$. The void volume PDF predicted by the CDE model deviates from the standard model (and the $\phi$CDM) prediction by $\sim10\%$ for large voids, a deviation that increases with void size. This deviation is not nearly as significant at the higher redshifts, nor is it seen between the $\phi$CDM and $\Lambda$CDM distributions across the three redshifts. This suggests that it is the coupling between the scalar field and dark matter that leaves a potentially observable imprint in the void abundance, particularly of large voids and at very late times, while the presence of the scalar field leaves only a minimal imprint. We propose that this signature may be attributed to the coupling, which introduces an additional drag force between baryons and dark matter, and causes thinning of void walls as matter evacuates more slowly from voids, enabling a greater number of larger voids to form through the merger of smaller voids \citep[building on the original suggestion by][]{2015MNRAS.446L...1S}. 

Additionally, we found that the distributions of void ellipticity and prolateness are extremely similar across all three models in the redshift range $z=0.0-1.0$. Our cosmological models leave no imprint on general void shape statistics, though it is possible that higher order shape statistics carry significant imprints. This indicates that general void shape is dominated by gravity and the initial phases of the density perturbations, and is insensitive to small changes in dark sector physics. General void shape statistics are therefore unlikely to be useful probes of cosmology, and more complex shape statistics that quantify irregularity may be necessary. Furthermore, we found that the distributions of ellipticity and prolateness, calculated through the fitting of ellipsoids to the true void shape, suggest that the most common ellipsoid that best fits the shape of each void has an ellipticity of $\approx0.10$ and a prolateness of $\approx-0.05$. Our results show that most voids are not, in fact, best represented by spheroids, but rather, they are best represented by slightly prolate ellipsoids. Furthermore, we find that the voids identified by our Hessian-based method can be very irregular in shape. Irregularity is most common amongst large voids, as they are likely to have become so through the merger of smaller voids. These results have important implications for common practices where the spherical void assumption is made (e.g. experiments involving void stacking, defining density profiles for voids). The inherent flaw in the assumption may significantly affect the results obtained through such practices. 

The differences in the distribution of average void densities among the cosmologies may be the most promising signature of underlying cosmology found in this study. With the exception of a great similarity between the average density PDFs in the $\phi$CDM and CDE models at $z=0.0$, all of the models predict distinctly different distributions for average void density, across all three redshifts. These results may be attributed to both the presence of the scalar field and its coupling to dark matter. CDE is seen to produce denser voids than $\phi$CDM, which suggests that the coupling alone affects rate of matter evacuation from the void. This may be attributed to the additional drag term experienced by baryons and dark matter, which would slow the movement of material out of the void. The fact that $\Lambda$CDM still produces the densest voids on average despite the effect of coupling in CDE suggests the effect of the scalar field in emptying voids is great enough to offset the effect of coupling. It is possible that the normalisation used in our simulations affected the position of the peaks derived from the $\phi$CDM and CDE simulations, and contributed to the strong similarity between those PDFs at $z=0.0$. Further studies involving the use of different normalisations need to be conducted to shed light onto this result. 

It is worth noting that the results presented in this study are true for specific realisations of quintessence and coupled dark energy cosmologies. It is not enough to simply determine that alternative cosmological models give rise to different void properties. As we have seen, different void properties may arise depending on the specific parameter values. Properly probing the nature of the dark sector requires quantifying how the strength of the coupling, or the specific changes in expansion history, affect the properties of voids. How strong must the coupling be to produce a deviation from the prediction of the standard model? Is there a dependence of the size of the deviation on the coupling strength? How different from $\Lambda$CDM predictions must the Hubble parameter be for any imprint to be observable? What form must the potential, $V(\phi)$, take in order to leave an imprint? These questions are especially important for constraining the vast range of possible models that are consistent with our universe. 
\section*{Acknowledgements}
E. A. acknowledges financial support by the Australian Government and The University of Sydney, through an Australian Postgraduate Award and a University of Sydney Merit Scholarship, respectively. P. J. E. acknowledges funding from the SSimPL programme and the Sydney Institute for Astronomy (SIfA), DP130100117 and DP140100198. The authors acknowledge the use of code written by Cullan Howlett to calculate non-linear power spectra, and the use of code written by Simeon Bird to read {\sc dark-gadget} simulation snapshots. The authors also acknowledge the University of Sydney HPC service at The University of Sydney for providing HPC resources, in particular the Artemis supercomputer, that have contributed to the research results reported within this paper. URL: \url{http://sydney.edu.au/research_support/}. This research made use of the NCI National Facility in Canberra, Australia, which is supported by the Australian Commonwealth Government, with resources provided by Intersect Australia Ltd and the Partnership Allocation Scheme of the Pawsey Supercomputing Centre.

\bibliographystyle{mnras}
\bibliography{Paper1} 








\bsp	
\label{lastpage}
\end{document}